\documentclass[conference]{IEEEtran}
\IEEEoverridecommandlockouts
\usepackage{cite}
\usepackage{amsmath,amssymb,amsfonts}
\usepackage{graphicx}
\usepackage{textcomp}
\usepackage{xcolor}
\usepackage{algorithm}
\usepackage{algpseudocode}
\def\BibTeX{{\rm B\kern-.05em{\sc i\kern-.025em b}\kern-.08em
    T\kern-.1667em\lower.7ex\hbox{E}\kern-.125emX}}
\begin{document}

\title{Singer Identification for Metaverse with Timbral and Middle-Level Perceptual Features\\
\thanks{$\ast$Corresponding author: Jianzong Wang (jzwang@188.com).}
}

\author{\IEEEauthorblockN{Xulong Zhang, Jianzong Wang$^{\ast}$, Ning Cheng, Jing Xiao}
\IEEEauthorblockA{\textit{Ping An Technology (Shenzhen) Co., Ltd., China} }
}


\maketitle

\begin{abstract}
Metaverse is an interactive world that combines reality and virtuality, where participants can be virtual avatars. Anyone can hold a concert in a virtual concert hall, and users can quickly identify the real singer behind the virtual idol through the singer identification. Most singer identification methods are processed using the frame-level features. However, expect the singer's timbre, the music frame includes music information, such as melodiousness, rhythm, and tonal. It means the music information is noise for using frame-level features to identify the singers. In this paper, instead of only the frame-level features, we propose to use another two features that address this problem. Middle-level feature, which represents the music's melodiousness, rhythmic stability, and tonal stability, and is able to capture the perceptual features of music. The timbre feature, which is used in speaker identification, represents the singers' voice features. Furthermore, we propose a convolutional recurrent neural network (CRNN) to combine three features for singer identification. The model firstly fuses the frame-level feature and timbre feature and then combines middle-level features to the mix features. In experiments, the proposed method achieves comparable performance on an average F1 score of 0.81 on the benchmark dataset of Artist20, which significantly improves related works.
\end{abstract}


\section{Introduction}
Metaverse~\cite{lee2021all} is about to take us into an epoch-making revolution. It will expand our existing physical world to a broader and boundless world that combines reality and virtuality. Metaverse can provide a more convenient and realistic work and entertainment experience. For example, a super large concert can be held in Metaverse, singers can use virtual avatars to perform super-imaginative performances, and can even use virtual singers to sing. Anyone with the dream of a singer can open his own concert in Metaverse, or it can be a parody show. For participating audiences, the singer identification technology can be used as a plug-in to quickly get the prompt of the singer information. Singer identification technology can also be used as a basis for imitation of the characters in the show to quickly switch.


Singer identification (SID) is an essential part of MIR, which purpose is to identify performing singers in a given audio sample \cite{ellis2007classifying,zhang2021singer}. SID is used in music library management to address the classification of songs by singers. Furthermore, the SID model is able to be used for downstream singing-related applications, such as similarity search, playlist generation, or song synthesis~\cite{lee2019learning,liu2019score,zhang2022MetaSID,humphrey2018introduction,aolan2021,gao2021vocal}.

In the preliminary studies, most scholars use traditional methods such as Gaussian Mixture Model (GMM), Hidden Markov Model (HMM), and the Support Vector Machine (SVM) to analyze the features in speech signals \cite{zhang2003automatic,zhang2022MDCNN-SID,mandel2006song,sibo2022,Dongaonkar,zhang2022TDASS}. However, the identification accuracy based on traditional techniques is not ideal. There are several possible explanations for the traditional methods shortages. On the one hand, the song's accompaniment is an intense noise for the SID task \cite{sturm2014simple}. In the SID model, eliminating the influence of accompaniment is an essential skill for improving the identification correct rate. But the traditional methods lack the ability. On the other hand, the difference in the singing voices of the two singers may not always be obvious. Because human beings share similar mechanisms in producing sounds, leading the similar voices \cite{hsieh2020addressing}. 

Most previous researches propose to separate singing from the accompaniment to address the accompaniment influence problem \cite{sha2013singing,mesaros2007singer,sharma2019importance,zhang2020research,qubo2021}. Su and Yang proposed the separation of phase information and sound \cite{su2013sparse}. The main idea is to separate and classify linguistic information, and the classification accuracy is 0.66. Open-unmix is an open-source tool with state-of-the-art performance in source separation to separate vocal and instrumental tracks of music\cite{stoter19,zhang2022SUSing}. It uses the data augmentation method and improves the accuracy of singer identification \cite{hsieh2020addressing}. 

To address human voice similarly, extracting features for robust representation are one of the important solutions. The previous methods, such as linear prediction coefficients (LPC), linear prediction cepstral coefficients(LPCC), and Mel spectrum cepstral coefficients(MFCC), mainly extract features related to speech recognition \cite{cai2011automatic,patil2012combining}. In addition, there are also some music recognition features, such as pitch, vibrato, \textit{etc.}, extracted to solve the problem \cite{nwe2007exploring}. The speaker timbre features are the distinguishing features used for identifying the speaker. Hamid \textit{et al.} propose to use i-vectors and timbre similarity to identify singers, which has significantly improved the accuracy \cite{eghbal2015vectors}. But i-vector is strongly influenced by the noise in the voice. Then, X-vector, which is obtained by a deep neural network (DNN), is proposed to reduce the noise influence for timbre features \cite{snyder2018x}. 

Except for the traditional methods, many scholars propose to use deep neural networks for SID. Nasrullah and Zhao \cite{nasrullah2019music} propose the convolutional recurrent neural network (CRNN) structure, which greatly improved the accuracy of singer identification. Zhang \textit{et al.} inspire on the principle of WaveNet~\cite{oord2016wavenet} vocoder and propose an end-to-end architecture in the time domain for singer identification. But whether or not the deep learning method is used in these models, most of them use the frame-level features to identify the singers. Frame-level features represent each frame of voice features, including the energy, \textit{etc.} It means the frame-level features are combined features of the singer's voices and the accompaniment. Even though the deep learning used in SID improves the accuracy, the accompaniment still influences the identified singer's results.

This paper proposes using two new types of features for singer identification to address the accompaniment influence. One is middle-level features which are obtained through the transfer learning of the music field. Middle-level features represent the stability of pitch and rhythm, the complexity of harmony and rhythm, \textit{etc.}, which can predict such as emotion, genre, and similarity. At the same time, it can capture music's perceptual features. We will give more detail and discuss the middle-level features in Section\ref{2.3}. The other is the timbre feature which is commonly used in speaker recognition. Among the different types of timbre features, we propose to use X-vector. Because it is more robust against noise. It means that even with the influence of accompaniment, X-vector can still distinguish different singers more accurately. We first propose a model using the frame-level audio features of spectrograms, middle-level features, and timbre features to make the singer's identification. To begin with, we use the convolutional network to fuse the combined feature from frame-level features and timbre features. Then the gated recurrent unit (GRU) is used to fuse the output features of the convolutional layer with the features obtained through the middle-level feature transfer learning. We used a public dataset, Artist20, to evaluate our model. The result shows it has improvement on the F1 score than the existing models.

Our main contributions are as follows:

\begin{itemize}
    \item We propose a new singer identification model, which uses the convolutional recurrent neural network (CRNN) and combines the frame-level audio features of spectrograms, middle-level features, and timbre features.
    \item We propose to use the middle-level features in the singer's identification task. Furthermore, we fuse the timbre features and frame-level to combined features, then combine them with middle-level features in the singers' identification task.
\end{itemize}


\section{Related Works}
In this section, we introduce the frame-level features, timbre features, and middle-level features. Besides, we introduce the previous model used in SID task.

\subsection{Frame-Level Audio Features} \label{2.1}

The music signal is also one type of voice signal with many characteristic parameters, such as frequency spectrum, linear prediction coefficient, linear prediction spectrum coefficient, etc. Mel-spectrum features and Mel-frequency cepstral coefficient (MFCC) are designed inspired by human ears, and it represents the non-linear relationship between frequency and pitch. They are widely used characteristic parameters in the field of audio identification. After discrete cosine transforms for log Mel-spectrum features to get the MFCC. They are results of Fast Fourier Transform for the original signal. However, because of the mathematical operation and wave filter used, some information is lost from the original signal. Some scholars propose directly input the original voice into the end-to-end neural network system to music information retrieval. Even though the end-to-end model resolves the information loss problem, the system required more learning data and increases time complexity. Therefore, we propose to use the Mel-spectrum features as frame-level audio features in the SID model.

\subsection{Timbral Features}
Timbre is different from different speakers. Therefore, the timbre feature was first applied to speaker recognition, more accurately distinguishing different speakers. X-vector\cite{snyder2018x} is one of the timbre features. Because of the statistics pooling layer in its network, X-vector can convert any length signal into fixed-length feature expressions. The X-vector model used a data augmentation strategy, including noise and reverberation in training. It makes the model more robust to interference such as noise and reverberation. Not only the speaker identification but also some other voice domain tasks, text-to-speech, voice conversion, \textit{etc.}, embedded X-vector for performance gains \cite{snyder2019speaker,pappagari2020x,jeancolas2021x,tang2022avqvc}.

\begin{table}[htbp]
  \centering
  \caption{Summary of middle-level features}
    \begin{tabular}{ccccc}
    \hline
     Perceptual Feature  &  mean  &  min  &  max  &  variance  \\
    \hline
    Melodiousness &          5.97  &          1.00  &        10.00  &          1.54  \\
    Articulation &          5.57  &          1.00  &          9.80  &          1.80  \\
    Rhythmic complexity &          4.66  &          1.00  &        10.00  &          1.44  \\
    Rhythmic stability &          6.30  &          1.20  &        10.00  &          1.47  \\
    Dissonance &          4.89  &          1.00  &          9.60  &          1.39  \\
    Tonal stability &          6.77  &          1.00  &        10.00  &          1.12  \\
    Modality &          5.48  &          1.60  &        10.00  &          1.35  \\
    \hline
    \end{tabular}%
  \label{tab:mlf}%
\end{table}%

\subsection{Middle-Level Features} \label{2.3}

The characteristics of the music field can be roughly divided into three levels. We call the clearly defined concepts such as beats and chords low-level features. Subjective concepts that are not clearly defined, such as mood, genre, and similarity, are called high-level characteristics. High-level characteristics can only be defined by considering all aspects of music. The ones in between are called middle-level features, such as the music's speed and the stability of rhythm. The concept of mid-level features was first proposed by Aljanaki and Soleymani \cite{aljanaki2018data} to improve music emotion recognition. They conclude seven mid-level musical features, including \emph{Melodiousness}, \emph{Articulation}, \emph{Rhythmic Complexity}, \emph{Rhythmic Stability}, \emph{Dissonance}, \emph{Tonal Stability}, and \emph{Modality}, each feature are continuous values between 1 and 10. They also provide audio and annotated example dataset of middle-level features. Besides, middle-level features have also been used for music retrieval and classification \cite{chowdhury2021towards}. Table \ref{tab:mlf} shows the descriptive statistical analysis of middle-level features.

\subsection{Singer Identification Models}
For traditional methods,  Ellis \textit{et al.} propose to use MFCC and Chroma-based features and Support Vector Machine (SVM) to identify singers \cite{ellis2007classifying}. SVM model aims to find a maximum-margin hyperplane divided by the target singer and others. Similarly, Su and Yang propose to base on SVM to use the codeword-based bag of features (BOF) \cite{DBLP:conf/ismir/SuY13}.  BOF means making a label for each frame of music to identify singers. Langlois \textit{et al}. propose to use Hidden Markov  Model (HMM) with MFCC features \cite{DBLP:conf/ismir/LangloisM09}.

For deep learning methods, Choi \textit{et al.} \cite{choi2017convolutional} prove that CNN could be used as a feature extractor for MIR tasks. At the same time, due to the characteristic that the sound is continuous along the time axis, the recurrent neural network (RNN) has also achieved great success in audio tasks such as speech recognition. Nasrullah and Zhao \cite{nasrullah2019music} propose a CRNN model for singers identification. The model consists of four CNN blocks, two GRU blocks, a fully connected layer, and a softmax function. CRNN combines the advantages of CNN and RNN, which lead the average F1 score on the album-split atrist20 dataset to achieve 0.603. Even though it is a significant improvement compared to the previous model, the model still does not fix the accompaniment's influence. To address it, Hsieh \textit{et al}. propose a data augmentation method to mix the different song's accompaniment and vocal \cite{9054069}. It means the singer's vocal will mix multiple accompaniments to be new songs. By using the CRNN model for augmented data, which is called CRNNM, the identify result achieves 0.73 on F1, which is the stat-of-the-art result in singer identification.

\section{Singer Identification Models}
In this section, we discuss our proposed model, including the features extraction, and architecture of the model.

\begin{figure}
  \centering
  \includegraphics[width=\linewidth]{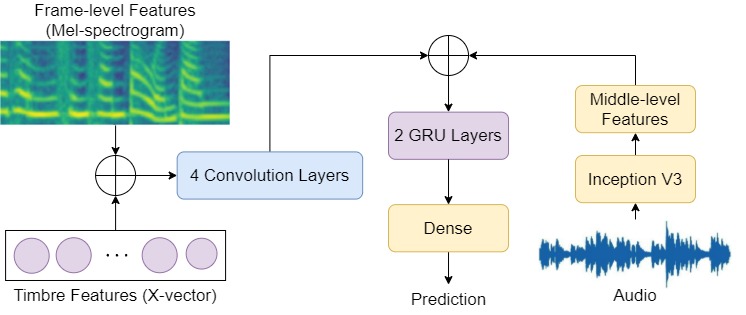}
  \caption{The architecture of the proposed SID model. After preprocessing, the input of SID is the original audio, frame-level features (Mel-spectrum), and timbre features (X-vector). To begin with, we use the convolution network to process the combined features from frame-level features and timbre features. Besides, by using the Inception V3 to extract the middle-level features from audio. In addition, the two GRU layers and dense to identify the singer through the combined features and middle-level features. The $\oplus$ symbol stands for channel-wise concatenation.}
  \label{fig:flow}
\end{figure}

\subsection{Features Extraction}
We aim to use the frame-level features, timbre features, and middle-level features to improve the performance of singer identification. The feature extraction algorithm is shown in Algorithm \ref{alg: feat extraction}.

\begin{algorithm}
\caption{Feature extraction of frame-level feature, timbre and the Middel-level feature. }\label{alg: feat extraction}
\textbf{Input:} audio waveform $x_{audio}$ \\
\textbf{Output:} fusion feature $y_{feat}$
\begin{algorithmic}[1]
\State  $feat_{mel}
\gets MEL(STFT(x_{audio}))$ \Comment{ get mel spectrum}
\State  $feat_{timbre} 
\gets TDNN(x_{audio})$ \Comment{ get timbre of x-vecotor}
\State $feat_{middle}^{L3},feat_{middle}^{L4},feat_{middle}^{L5} 
\gets InspectionV3(x_{audio})$ \Comment{ get middle feature}
\State $fusion_{concat} 
\gets Stack(feat_{mel},feat_{timbre})$
\State $fusion_{cnn} 
\gets CNN(fusion_{concat})$
\State $y_{feat} 
\gets Concat(fusion_{cnn},feat_{middle}^{layer}), layer \in \{L3,L4,L5\}$
\end{algorithmic}
\end{algorithm}

\subsubsection{Extract of Frame-Level Features}

We propose to use the mel-spectrogram as the frame-level feature. As we discussed in Section \ref{2.1}, we first make Short-term Fourier Transform (STFT) for the original signal to get the spectrogram information. By using a mel filter for the spectrogram, we get the mel-spectrogram.
The STFT equation is shown as below,

\begin{align}
feat_{spec} = \sum\limits_{n = - \infty }^\infty x [n]w[n - m]{e^{ - j\omega n}}
\label{eq1}
\end{align}
where, $x[n]$ represents the input signal and $w[n]$ describes the window function, such as Hamming window.

 Then we transfer it to the mels domain as the following equation.

\begin{equation}
    feat_{mel} = 2595log_{10}(1+\frac{feat_{spec}}{700})
\end{equation}
\noindent where, the $feat_{mel}$ represents the mel-spectogram features, $feat_{spec}$ is the spectrogram features same in Equation \ref{eq1}.

\subsubsection{Extract of Timbre Features}

We propose to use the X-vector as the timbre feature. To extract the features, we used the Snyder \textit{et al.}\cite{snyder2018x} proposed Deep Neural Network (DNN) before the singer identification. The DNN model consists of multiple TDNN layers, a statistics pooling layer, two embedding layers, and a Softmax layer. TDNN layer is also a multiple layer neural network to be able to learn the sequence information. The statistics pooling layer is used to combine each frame to represent the input voice features. In our model, X-vector is a 512 dimensions vector.

\subsubsection{Extract of Middle-Level Features}

The middle-level features used in this paper are obtained through the transfer learning of the original middle-level features of music~\cite{aljanaki2018data}. This feature can reflect the middle-level features of music such as Melodiousness, Rhythmic stability, and Rhythmic complexity, and can capture Perceptual features of music perception.

The middle-level feature network is the same as Inception V3 used in the previous work~\cite{aljanaki2018data}. We input the singing voice into the model to obtain the middle-level features. But instead of using the final output of the model as the middle-level features, we extracted the outputs of the third, fourth, and fifth layers of the Inccriteria V3 network to represent the features. We called the L3, L4, and L5 middle-level features for different layers result. Such operations are also compatible with transfer learning. The three middle-level features are as follows:

\textbf{L3 middle-level features}: Represents the inverse third level of Incetption V3 as a representation of the middle-level feature, which is 1-by-128 dimensions.

\textbf{L4 middle-level features}: Represents the inverse fourth level of Incetption V3 as a representation of the middle-level feature, which is 1-by-128 dimensions.

\textbf{L5 middle-level features}: Represents the inverse fifth level of Incetption V3 as a representation of the middle-level feature, which is 1-by-256 dimensions.

\subsection{CRNN with Multiple Features}

For using the multiple features, we propose the SID model shown in Figure \ref{fig:flow}. We preprocess the frame-level features and timbre features from audio to get the mel-spectrogram and X-vector.

To begin with, we combine the mel-spectrum features and X-vector features to combined features. Besides, to use inspection V3 \cite{szegedy2016rethinking} to extract the middle-level features. In addition, to use the four-layer convolutional layers to fuse the combined features. The convolutional layers aim to learn the critical information of the combined features. Then, we combine the output of convolutional layers and middle-level features. Furthermore, the two-layers GRU \cite{DBLP:journals/access/ChengSCLCCH21} is used to learn the new combined features. GRU layer aims to find the sequence information of the music. Finally, we use the ultimate fusion features to identify singers through the dense layer.



\section{Experiments}
The models are evaluated using Artist20 under the album split, averaging the F1 scores of three independent runs.

\subsection{Dataset and Audio Processing}
Artist20, a music artist identification dataset created by Ellis \cite{ellis2007classifying}, was used in this paper to evaluate classification performance. It contains six albums by 20 artists, covering a variety of musical styles.The summary is given in Table \ref{tab:Artist20}.


\begin{table}[htbp]
  \centering
  \caption{Artist20 dataset specifications}
    \begin{tabular}{cc|cc}
    \hline
    \textbf{Property} & \textbf{Value} & \textbf{Property} & \textbf{Value} \\
    \hline
    Bitrate & 32 kbps & Channels & Mono \\
    Sample Rate & 16 kHz & Albums Per Artist & 6\\
    \hline
    Total Tracks & 1413 & Total Artists & 20 \\

    \hline
    \end{tabular}%
  \label{tab:Artist20}%
\end{table}%

The Artist20 dataset is the most widely used dataset for artist identification. It is composed of 20 artists; each artist has 6 albums and a total of 1413 songs. In the literature of SID, data splitting can be done in two ways: song-split or album-split. The former split a dataset by randomly assigning songs to the three subsets. In contrast, the latter makes sure that songs from the same album are either in the training, validation or the test split. According to the deep learning model principle, splitting the dataset into training and testing is an important experiment setting. For singer identification, a song always has the same part, especially in the refrain. If we randomly split the song, these same frames may be arranged into both training and testing datasets.  Besides, Whitman \textit{et al}. \cite{whitman2001artist} finds the producer effect. It means the songs in the same albums have a similar style. The style is useless for singer identification as a singer always has multiple styles of songs. Suppose split the albums' songs to training and validating datasets, the learned features in the album's style, not the target singer. According to the two special reasons in singer classification, splitting the dataset by album is the better choice to avoid the noise from the same frames in the songs and the same style. Therefore, the accuracy for song-split might be overly optimistic and tends to be higher than that of album-split. In the evaluation, we focus on and only consider album-split in our work.

To follow the album split, we divided four of the six albums into training sets, and the remaining two were divided into validation sets and test sets. We also chunk the data, cutting each song according to a window length of 30 seconds and a sliding time of 10 seconds.

\subsection{Experimental Setting}

We set the SVM (SFCC features) \cite{ellis2007classifying}, HMM\cite{DBLP:conf/ismir/LangloisM09}, SVM\cite{DBLP:conf/ismir/LangloisM09}, Nerueal Network\cite{7489950}, CRNN model \cite{nasrullah2019music}, and state-of-the-art CRNNM model \cite{9054069} as the baseline, to compare with our proposed model. The propose model uses frame-level features, mel-spectrogram, to identify the singers. It is called CRNN+X-vector+middle-level features that show the different features used.

Furthermore, to analyze each feature contribution in the singer identification, we set an ablation experiment for different features used. Except for the previous model and the proposed one, we add the CRNN+X-vector model which uses frame-level features and timbre features. 

Moreover, we set the second ablation experiment to determine which middle-level feature is more suitable to identify the singer. There are three comparable models which use one of the L3, L4, and L5 middle-level features. The model is called as CRNN+X-vector+\emph{\{middle-level feature\}}, middle-level feature $\in$ [L3, L4, L5].

Besides, we set the third ablation experiment. We purpose to study the combined different layers of middle-level features, such as combined L3 and L4, etc, how to influence the identification result. We set the model name as CRNN+X-vector+\emph{\{used middle-level features\}}. But we set CRNN+X-vector+All to represent the model that uses L3, L4, and L5 together.



\begin{table}[htbp]
  \centering
  \caption{Average testing F1 score on the Artist20 dataset.}
    \begin{tabular}{cccc}
    \hline
    Type & Model & F1/best & F1/avg \\
    \hline
    Baseline & SVM (MFCC  features)\cite{ellis2007classifying}  & 0.49 & 0.48  \\
    Baseline & HMM\cite{DBLP:conf/ismir/LangloisM09}  & 0.52 & 0.51  \\
    Baseline & SVM (BOF features)\cite{DBLP:conf/ismir/SuY13}  & 0.58 & 0.56  \\
    Baseline & Neural Network\cite{7489950}  & 0.57 & 0.56  \\
    Baseline & CRNN\cite{nasrullah2019music}  & 0.61 & 0.60  \\
    Baseline & CRNNM\cite{9054069}  & 0.75 & 0.73  \\
    \textbf{Our} & \textbf{CRNN+X-vector+L4} &   \textbf{0.86} & \textbf{0.81}   \\
    \hline
    \end{tabular}%
  \label{tab:addlabel}%
\end{table}%

\begin{table}[htbp]
  \centering
  \caption{Average testing F1 score for ablation experiments. AT\uppercase\expandafter{\romannumeral1} to AT\uppercase\expandafter{\romannumeral3} represents the ablation experiment \uppercase\expandafter{\romannumeral1} to \uppercase\expandafter{\romannumeral3}.}
  \begin{tabular}{cccc}
  \hline
    Type & Model & F1/best & F1/avg \\
  \hline
    AT\uppercase\expandafter{\romannumeral1} & CRNN+X-vector &  0.72 & 0.71    \\
    AT\uppercase\expandafter{\romannumeral2} & CRNN+X-vector+L3 &   0.81 & 0.80   \\
    \textbf{AT\uppercase\expandafter{\romannumeral2}} & \textbf{CRNN+X-vector+L4} &   \textbf{0.86} & \textbf{0.81}   \\
    AT\uppercase\expandafter{\romannumeral2} & CRNN+X-vector+L5 &   0.79 & 0.78   \\
    AT\uppercase\expandafter{\romannumeral3} & CRNN+X-vector+L3+L4 &   0.71 & 0.70   \\
    AT\uppercase\expandafter{\romannumeral3} & CRNN+X-vector+L3+L5 &   0.80 & 0.80   \\
    AT\uppercase\expandafter{\romannumeral3} & CRNN+X-vector+L4+L5 &   0.77 & 0.76   \\
    AT\uppercase\expandafter{\romannumeral3} & CRNN+X-vector+All &   0.78 & 0.77   \\
    
    \hline
    \end{tabular}%
  \label{tab:AT}%
\end{table}%

\subsection{Results and Discussion}
\textbf{Overall Comparisons.} Table \ref{tab:addlabel} shows the result of identify singers. The single most striking observation to emerge from the data comparison was the \emph{CRNN+X-vector+L4} model achieves 0.86 in terms of F1 score, which improves 0.1 than the stat-of-the-art model. Our proposed model obtains the best result and addresses the accompaniment influence without data augmentation. 

\textbf{Ablation experiment \uppercase\expandafter{\romannumeral1}}. As can be seen from Table\ref{tab:AT}, when embedding the X-vector features to CRNN, the result improved 0.11 F1 scores. It means the timbre features are beneficial for identifying singers. But the F1 score for CRNN+X-vector is smaller than the proposed model. The reasonable explanation is middle-level features introduced in this paper are positively correlated to the identification.

\textbf{Ablation experiment \uppercase\expandafter{\romannumeral2}}. The more surprising finding is that the L5 middle-level features reduce the F1 score from 0.86 to 0.79. A possible explanation for this might be that the deeper the inception V3 model, the more fuse operation in middle-level features, leading the over-fitting. It means the model may be wrong to believe some identifying information is noise and remove it. Because the inception V3 is a multiple convolution layers model, it will lose information during each operation. The possible explanation for L3 worse than L4 is that the convolution of the third layer of inception V3 is under-fitting. It means the convolution result in the third layer still has noise in it.

\textbf{Ablation experiment \uppercase\expandafter{\romannumeral3}}. What is surprising is that combining more than one middle-features is not helpful for identifying singers and reduces the result. This result may be explained by the fact that L3, L4, and L5 middle-features have learned similar contents and have multicollinearity. The multicollinearity affects the CRNN and gets a bad result.

In summary, these results show that the timbre features and middle-level features are important for singer identification and useful for reducing the accompaniment influence. This is a significant result that our proposed model achieves 0.86 F1 scores on Artist20, which is higher 0.2 than the baseline. 

\begin{figure}[htb]
  \centering
  \includegraphics[width=\linewidth]{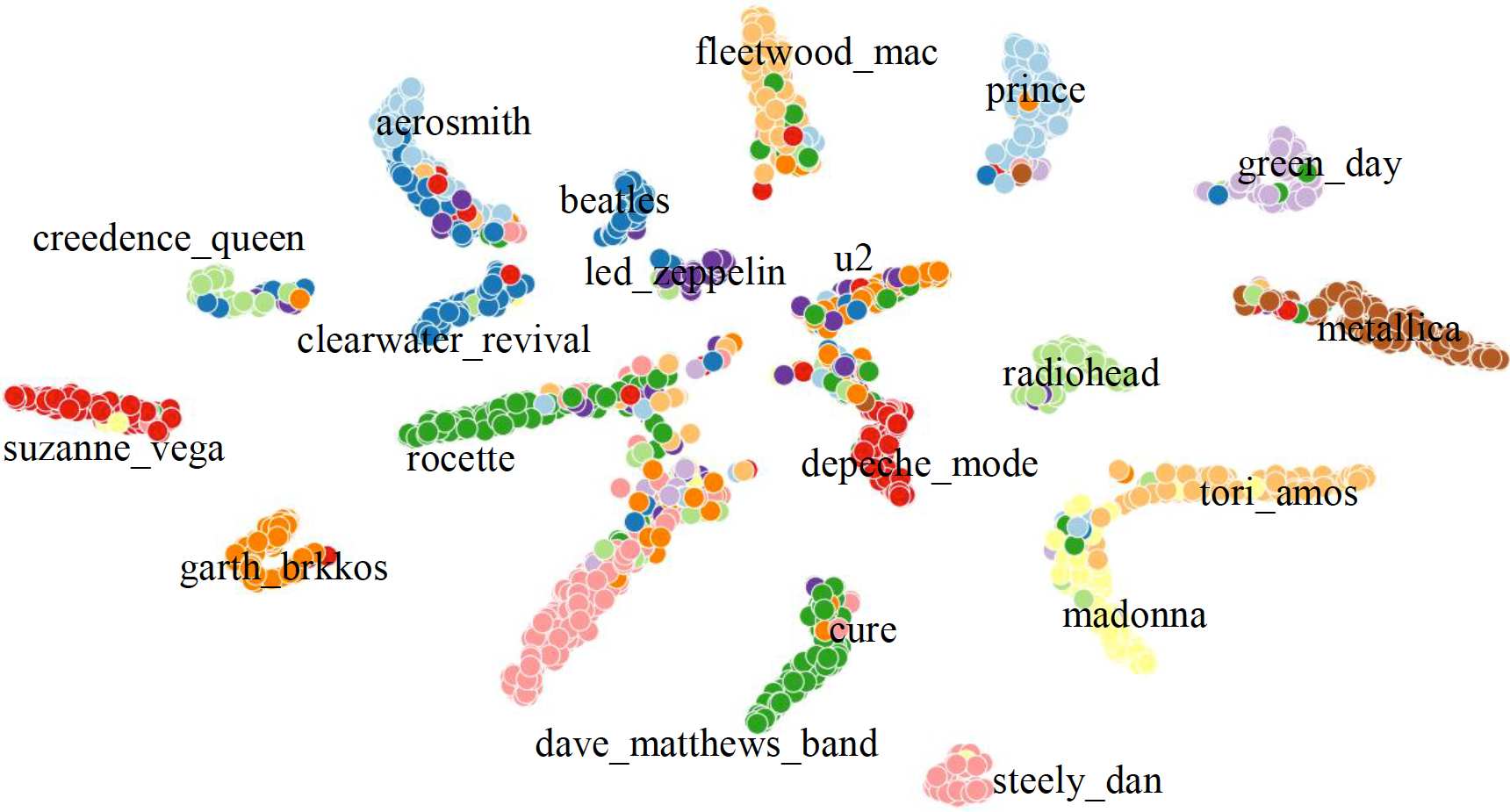}
  \caption{The t-SNE Visualization of \emph{CRNN+X-vector+L4 middle-level} model. Each color represents a class, the number of classes is 20.}
  \label{fig:speech_production}
\end{figure}

\subsection{Visualization}
To show our model performance, this paper uses an extra method than only the F1 scores. The visualization method is able to show the different classes and their distance between each class in a graph. This paper uses t-distribution random neighbor embedding (t-SNE)\cite{van2008visualizing} to do the visualization. The t-SNE is able to reduce the dimension of data, and the classification can be explored intuitively. At the bottleneck layer of the network shown in Figure \ref{fig:flow}, to directly carry out the final full connection layer, each audio sample is transformed into a vector for classification. These vectors describe the representations that the model learns to use for segmentation.  

Figure \ref{fig:speech_production} shows the visualization of the singer identification result by \emph{CRNN+X-vector+L4} model. It proves singers are successfully divided into 20 categories, and the distance is far apart from each other. It is a significant improvement from the previous CRNN\cite{nasrullah2019music}.

\section{Conclusions}
For singer identification in the Metaverse, we introduce two new features: the timbre feature and the features obtained through the transfer learning of the middle-level features of music. Then, the convolutional network and recurrent network were used to fuse Mel-spectrum, timbre, and features obtained through middle-level feature transfer learning and then use the final fusion features to identify singers. The evaluation results on a benchmark dataset of Artist20 show that the best performing model achieves an average F1 score of 0.86 across three independent trials, which is a substantial improvement over the corresponding baseline under similar conditions. Furthermore, the empirical results show that both the timbre features and the obtained through the transfer learning of the middle-level features of music improve performance, especially the latter. 
The features introduced in this article can be applied to other singing voice-related tasks such as music genre classification. 

\section{Acknowledgement}
This paper is supported by the Key Research and Development Program of Guangdong Province under grant No.2021B0101400003. Corresponding author is Jianzong Wang from Ping An Technology (Shenzhen) Co., Ltd (jzwang@188.com).


\bibliographystyle{IEEEtran}
\bibliography{mybib}

\end{document}